\documentclass[twocolumn,secnumarabic,amssymb, nobibnotes, aps]{revtex4-2}

\usepackage{graphicx}

\begin{document}
	
\title{Optical trapping and manipulation of fluorescent polymer-based nanostructures: measuring optical properties of materials in the nanoscale range}%
	
\author{T. A. Moura$^{1,2}$}
\author{M. L. Lana Júnior$^{1}$}
\author{C. H. V. da Silva$^{1}$}
\author{L. R. Américo$^{1}$}
\author{J. B. S. Mendes$^{1}$}
\author{M. C. N. P. Brandão$^{1}$}
\author{A. G. S. Subtil$^{1}$}
\author{M. S. Rocha$^{1}$}
\email{marcios.rocha@ufv.br}
\affiliation{$^{1}$Departamento de F\'isica, Universidade Federal de Vi\c{c}osa. Vi\c{c}osa, Minas Gerais, Brazil.}%
\affiliation{$^{2}$Brazilian Nanotechnology National Laboratory (LNNano), Brazilian Center for Research in Energy and Materials (CNPEM). CEP 13083-100. Campinas, São Paulo, Brazil.}%

\date{\today}%
	
	
\begin{abstract}

We present a novel approach to determine the optical properties of materials in the nanoscale range using optical tweezers (OT). Fluorescent polymer-based nanostructures (pdots) are optically trapped in a Gaussian beam OT and the trap stiffness is studied as a function of various parameters of interest. We explicitly show that properties such as the refractive index and the optical anisotropy of these nanostructures can be determined with high accuracy by comparing the experimental data to an optical force model. In particular, we demonstrate that the effective optical properties of these pdots can be modulated by changing the light wavelength that excites the sample, opening the door for a fine tuning of their optical response, with possible applications in the development of new sensors and/or other optoelectronic devices.

\end{abstract}
	
\keywords{optical tweezers, pdots, nanophotonics}
\maketitle

The optical trapping and manipulation of particles made of new (``unusual'') materials is an emerging field with a concrete potential to generate the next revolution in the optical tweezers (OT) community and correlated areas, because the wide range of properties of these materials allow many novel possibilities \cite{Spesyvtseva}, from the application of femto-Newton forces with high accuracy \cite{MouraSi} to the development of devices such as microrheometers and single molecule thermal machines \cite{Campos, Campos2, Campos3}. Furthermore, the optical trapping and manipulation of nanostructures such as quantum dots, nanotubes, graphene and others was also recently demonstrated \cite{MaragoNatNano, SudhakarGe, Kolbow}, opening the door to use OT as an important tool in fundamental nanotechnology studies, as well as in the innovative development of nanodevices.

In the present work we extend the use of optical tweezers for characterizing the optical properties of materials in the nanoscale range. To develop this approach, we studied the optical trapping and manipulation of a new type of material for the very first time: fluorescent polymer nanostructures (pdots); in this case, poly[2-methoxy-5-(2'-ethylhexoxy)-p-phenylene vinylene] (MEH-PPV) pdots. Our method in fact allows one to measure the optical properties of individual trapped pdots, being an approach that can be applied to determine quantities such as the refractive index and the optical anisotropy of these nanostructures. Furthermore, we show that these optical properties can be modulated by changing the excitation wavelength that illuminates the sample, a feature important to the development of new sensors and/or other optoelectronic devices. Usually, such properties are determined for much larger samples like the bulk material itself or thin films. Therefore, the present work opens the possibility to study, for example, how the size of the nanostructures affects their effective optical response. To the best of our knowledge, this is the first approach that allows one to measure the optical properties of individual isolated nanostructures.

\begin{figure*}
	\centering
	\includegraphics[width=15cm]{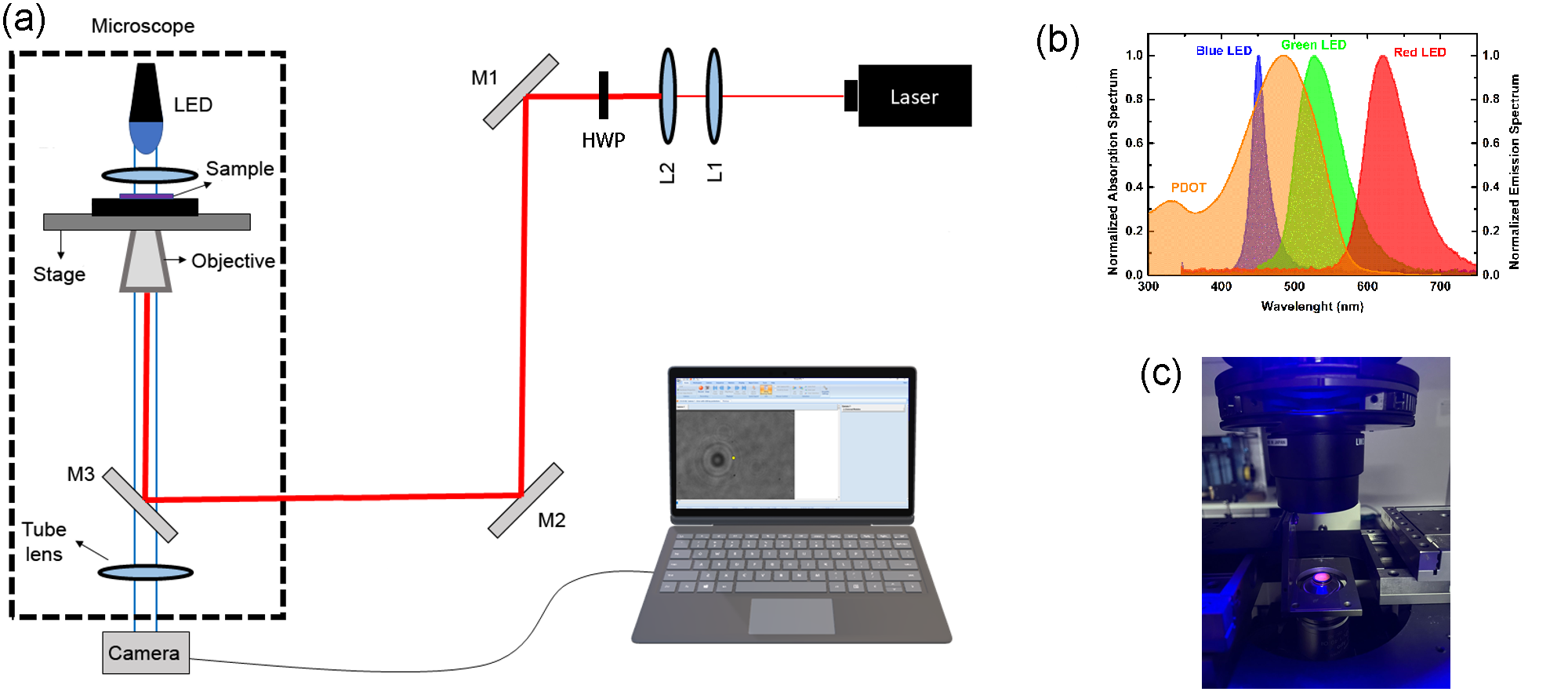}
	\caption{(a) Schematic drawing of the experimental setup used in our optical tweezers. L1 and L2 are lenses used to expand the beam diameter. M1, M2 and M3 are mirrors that control the beam propagation direction and HWP is a half-wave plate to control the laser polarization direction. (b) Emission spectra of each LED used to illuminate the sample and the absorption spectrum of the pdots. (c) A photography of the real setup using the blue illumination LED, where one can note the pdots fluorescing in the sample chamber.}
	\label{setup}
\end{figure*}

The OT setup used in this work consists of a linearly polarized near-infrared ($\lambda_0$ = 1,064 nm) solid state Gaussian (TEM$_{00}$) laser beam (CNI Laser, China), mounted in an inverted microscope (Nikon Ti-S) with a 100$\times$ NA 1.4 objective. The original illumination system of the microscope was substituted by a LED illumination system with selectable wavelengths in the visible range. Here we used three distinct wavelengths to excite the pdots trapped in the tweezers in order to study the effect of such excitation in their optical properties, which is reflected in the trapping of these nanostructures. Finally, the optical properties in each situation are determined by comparing the experimental data with a theoretical model for the optical force. In Figure \ref{setup} we show a schematic drawing of the OT setup (panel \textit{a}) and the emission spectra of each LED along with the absorption spectrum of the pdots (panel \textit{b}). Observe that the pdots absorption in the laser wavelength used for the OT (1,064 nm) is negligible.
A photography of the real setup using the blue illumination LED is also shown for reference, where one can note the pdots fluorescing in the sample chamber (panel \textit{c}). The LED system was carefully aligned and adjusted such that the sample is subjected to the same photon flux independent on the wavelength selected. This fact guarantees that the effects that will be presented here are exclusively due to the different wavelengths used, and not due to a change in the photon flux reaching the samples. 

The experiments consist in measuring the trap stiffness of the system (OT + trapped pdots in water) varying distinct parameters of interest, with the goal to characterize the trapping of such nanostructures under various conditions, specially when they are excited by different wavelengths. The methodology employed to perform these measurements is described in the Supplemental Material. 

In Fig. \ref{kxky} (panels \textit{a} and \textit{b}) we show the behavior of the transverse trap stiffness measured along two perpendicular directions ($\kappa_x$, $\kappa_y$) of a 365 nm radius pdot (at 10 $\mu$m above the bottom of the sample chamber) as a function of the laser power at the focus, for the three different illumination wavelengths. In panels \textit{c} and \textit{d} the same quantities are shown for different pdots as a function of their radii, for a fixed laser power of 25 mW at the focus. Finally, in panel \textit{e} we show the ratio $\kappa_x$/$\kappa_y$ as a function of the pdots radius. 
	
\begin{figure*}
	\centering
	\includegraphics[width=12cm]{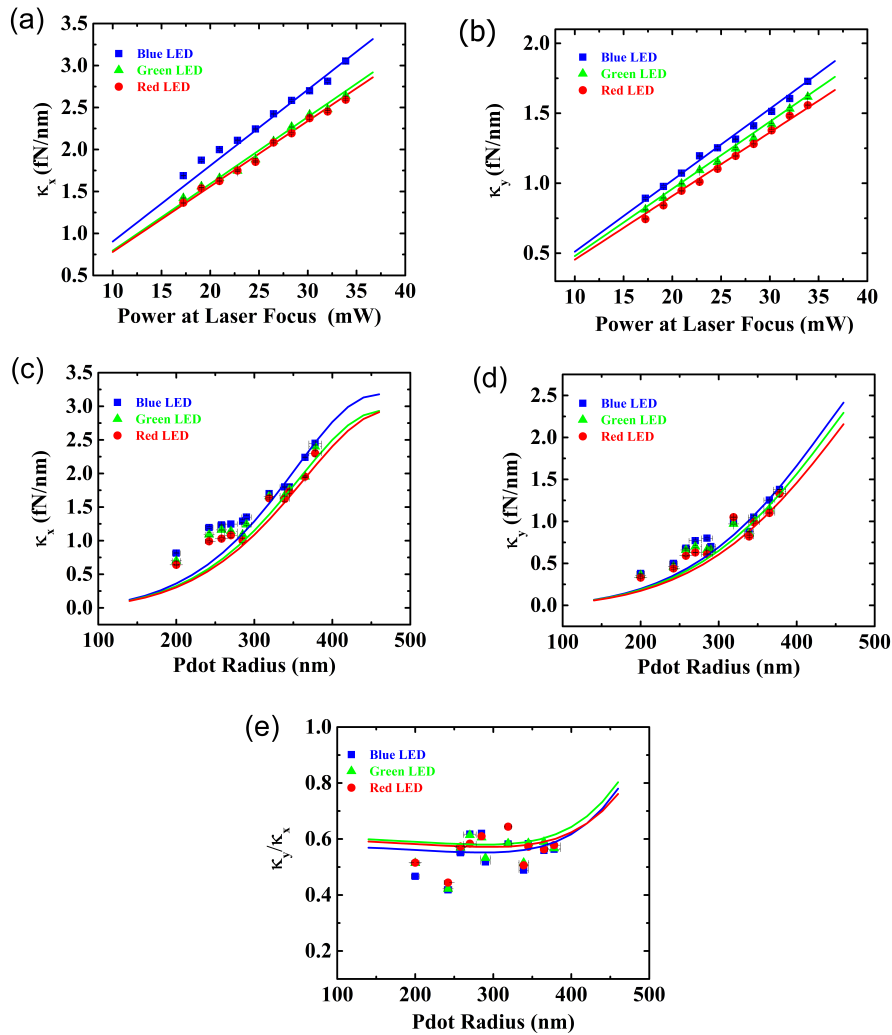}
	\caption{(\textit{a}) and (\textit{b}) Transverse trap stiffness measured along two perpendicular directions ($\kappa_x$, $\kappa_y$) for a 365 nm radius pdot (at 10 $\mu$m above the bottom of the sample chamber) as a function of the laser power at the focus, for the three different illumination wavelengths. (\textit{c}) and (\textit{d}): same quantities for different pdots as a function of their radius, for a fixed laser power of 25 mW at the focus.  (\textit{e}) $\kappa_x$/$\kappa_y$ as a function of the pdots radius. The error bars shown are the calculated standard error from a series of independent measurements.}
	\label{kxky}
\end{figure*}

The solid lines in all panels of Fig. \ref{kxky} are the predictions of a simple optical force model based in the Rayleigh regime, valid for particles smaller than the laser wavelength employed, which is the case here \cite{RochaAJP2}. Such model is described in detail in the Supplemental Material. Observe that the theory agrees very well with the experimental data in each case, showing that the proposed model works well for the current situation. By comparing the theoretical model with the experimental data, we can determine the best values of the fitting parameters that generate the solid curves shown in Fig. \ref{kxky}. For the current proposed model these parameters are the refractive index of the nanoparticles $n$ (at 1,064 nm, the wavelength used in the trapping laser) and the polarization asymmetry parameter $\eta$, previously defined in ref. \cite{Campos3} (see also the Supplemental Material). All other parameters needed to generate the theoretical curves of Fig. \ref{kxky} were measured for our setup; they are the beam waist at the equilibrium trapping position $\omega_e$ = (0.99 $\pm$ 0.09) $\mu$m, the laser power at the focus and the particle radius itself (see the Supplemental Material). Such an approach can thus be used to determine unknown optical properties of the nanostructures, as anticipated. 

\begin{figure*}
	\centering
	\includegraphics[width=12cm]{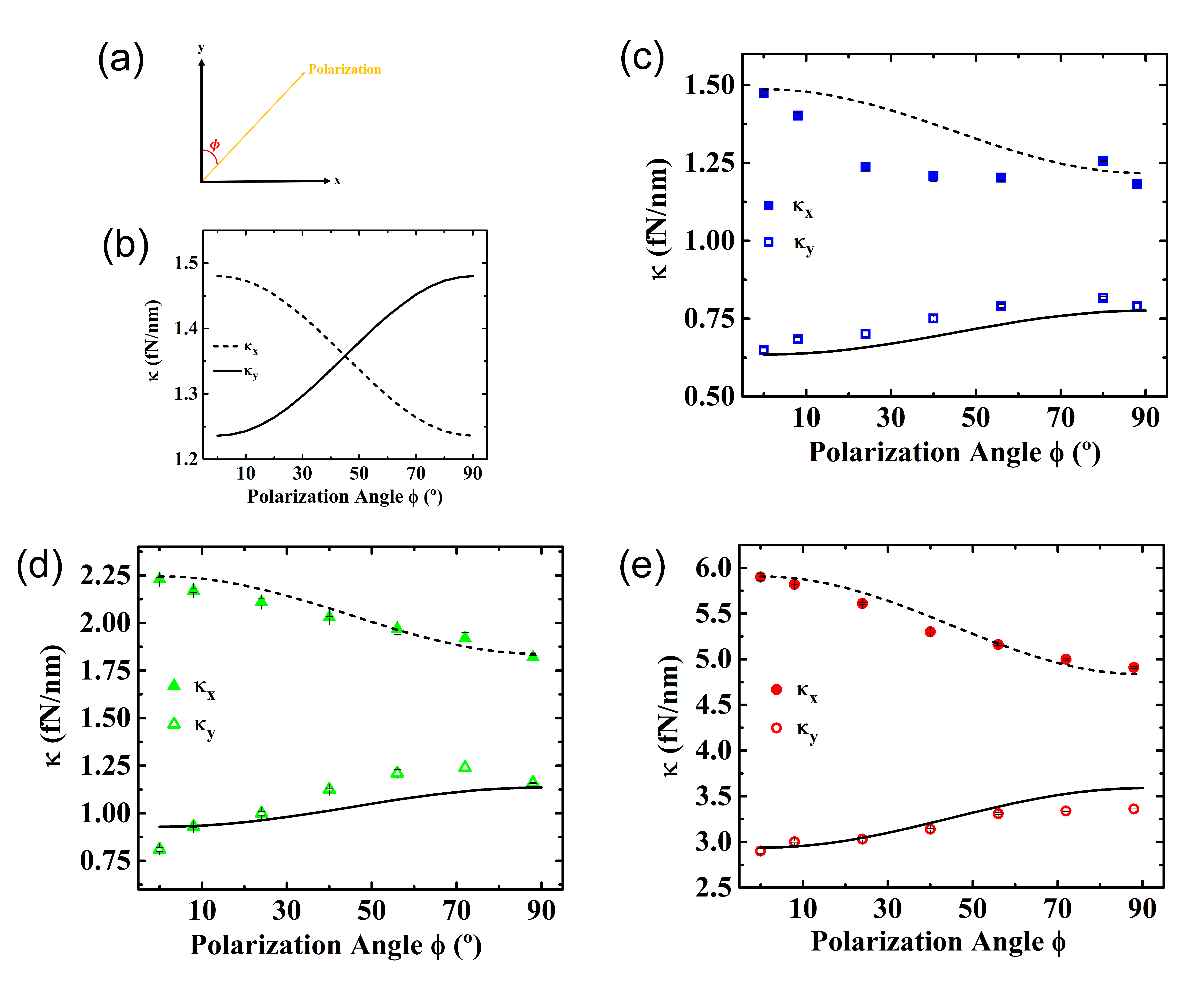}
	\caption{(\textit{a}) The polarization direction of the laser reaching the microscope, which forms the azimuthal angle $\phi$ with the $y$-axis. (\textit{b}) Theoretical prediction from our model for the the trap stiffness ($\kappa_x$, $\kappa_y$) as a function of $\phi$ without anisotropy. (\textit{c}), (\textit{d}) and (\textit{e}) Measured behavior of $\kappa_x$ and $\kappa_y$ for the three used illumination LEDs, along with the theoretical predictions of the model obtained using the refractive indices and asymmetry parameters show in Table \ref{tableresults}. Panel (\textit{c}): 315 nm radius particle; panel (\textit{d}) 295 nm radius particle; panel (\textit{e}) 440 nm radius particle. All measurements were performed using a laser power of 45 mW at the focus and with the particles distant 10 $\mu$m from the bottom of the sample chamber. The error bars shown are the calculated standard error from a series of independent measurements.}
	\label{anisotropy}
\end{figure*}

Interestingly, for our pdots the fitting parameters strongly depend on the LED wavelength used to illuminate the sample, suggesting that partial light absorption by the particles at these wavelengths changes their optical properties, reflecting in different trap stiffness values measured at each wavelength. This result is in agreement with previous reports for the MEH-PPV material. Van der Horst \textit{et al.} \cite{vanderHorstExciton} and Warman \textit{et al.} \cite{WarmanExciton} have in fact show that the polarizatility of this polymer increases at the excited state, a change related to the generation of excitons. In fact, the majority of semiconducting polymers exhibit a wide direct band gap and can present intense fluorescence emission. Conjugated polymers have delocalized $\pi$-electrons along the chain, and transitions occur from the ground to the excited state when they absorb light. Studies show that the emission processes of MEH-PPV in solution depend strongly on its interaction with the environment. The emission spectra is determined by the nature of the emitting species, which can originate from both an excitonic state or be attributed to aggregates. Nevertheless, the MEH-PPV exhibits a collapsed state (pdots) when in an aqueous environment, which enhances interchain interactions resulting in predominantly excitonic emission characteristics \cite{ZhengExciton, GhoshExciton}. The above discussion explains the variation measured in the refractive index of the pdots when the illumination wavelength is changed. Furthermore, panels (\textit{a}) and (\textit{b}) of Fig. \ref{kxky} show a consistent linear behavior of the trap stiffness as a function of the laser power. Such a behavior confirms that the generation of excitons is the main mechanism behind polarizability changes, because the other possibility (electron-hole generation) induces deviations from the linearity for this type of data at moderate to high laser powers, a phenomenon related to the variation of the effective absorption coefficient due to free charge carriers \cite{Andrade, OliveiraPANI}. In addition, the refractive indices obtained from the $\kappa_x$ and $\kappa_y$ data at each wavelength, $n_x$ and $n_y$, are different in all cases, suggesting that the pdot particles present anisotropy, responding differently in two perpendicular directions - a result that will be confirmed in the data of Fig. \ref{anisotropy}. In Table \ref{tableresults} we show the fitting parameter results obtained for the refractive indices and polarization asymmetry parameters at the three LED wavelengths used.

\begin{table}[h]
	\centering \caption{Fitting parameter results obtained for the refractive indices and asymmetry parameters at the three LED wavelengths used. The error bars estimated from the fittings are $<$ 0.007$\%$ for the refractive indices and $<$ 2.5$\%$ for the asymmetry parameters.}
	\label{tableresults}
	\begin{tabular*}{\columnwidth }{@{\extracolsep{\fill} } cccc}
		\hline\hline\noalign{\smallskip} LED & $n_x$ & $n_y$ & $\eta$    \\
		\hline \noalign{\smallskip} Blue  & 1.428 & 1.382 &  0.03    \\
		\noalign{\smallskip}   Green  & 1.423 & 1.379 &  0.07  \\
		\noalign{\smallskip}   Red  & 1.421 & 1.376 & 0.07   \\
		\hline\hline \noalign{\smallskip}
	\end{tabular*}
\end{table}

The fact that the trap stiffness is higher at smaller wavelengths indicate that light absorption by the particle, which is larger for the blue and green LEDs (see Fig. \ref{setup}$b$), changes their optical properties deepening the optical potential well that characterize the gradient force of the OT on the particles. Table \ref{tableresults} results explicitly demonstrate this fact, with the refractive index increasing for smaller wavelengths. Therefore, our experiments explicitly show that the effective optical properties of our pdots can be easily modulated simply by changing the illumination wavelength, suggesting that these properties can be suitably tuned for specific purposes, for example, in the development of new sensors or other photonic devices. Furthermore, it is worth to note that our methodology is a very accurate method to determine the refractive index (and possible other optical properties in the future, depending on the force model used) of nanostructures, an issue that is hard to be done with traditional bulk techniques. 

The other interesting result previously mentioned is the fact that the measured trap stiffness along the two perpendicular directions ($\kappa_x$, $\kappa_y$) are systematically different for the three wavelengths tested, indicating that the pdots present a considerable anisotropy, with optical responses that depend on the angle between the laser polarization direction and the given studied direction ($x$ and $y$ in the case). Tilley \textit{et al.} reported the anisotropic character of MEH-PPV polymers using fluorescence anisotropy measurements \cite{TilleyAnisot}. In addition, previous works investigated the optical trapping and manipulation of anisotropic particles of various materials, presenting calculations for the optical forces \cite{SimpsonAnisotropy, DraineAnisotropy, SimpsonAnisotropy2} and torques \cite{LaPortaAnisotropy, SimpsonAnisotropy2}, but the news here concerns the trapping of polymer-based nanostrucures and the determination their optical properties  from trap stiffness measurements. 

Here the anisotropy of the MEH-PPV pdots was investigated further by changing the  laser polarization direction (measured when incident on the objective back aperture), which is defined by the azimuthal angle $\phi$ measured in relation to the $y$-axis, as depicted in Fig. \ref{anisotropy}(\textit{a}). In the same figure, panel (\textit{b}), we show the theoretical prediction of our model for the trap stiffness ($\kappa_x$, $\kappa_y$) of a 365 nm radius pdot as a function of $\phi$ without anisotropy, \textit{i. e.}, considering that the refractive indices corresponding to the two perpendicular directions $x$, $y$ are equal (to generate the figure, we used 1.423 for these indices). Such result is qualitatively identical to those obtained experimentally by So and Choi using polystyrene beads, which are isotropic \cite{SoChoi}.  In panels (\textit{c}), (\textit{d}) and (\textit{e}) the measured behavior of $\kappa_x$ and $\kappa_y$ are show for the three used illumination LEDs, along with the theoretical predictions of the model obtained using the refractive indices and asymmetry parameters show in Table \ref{tableresults}. Observe that the anisotropy introduces a fundamental difference on the behavior of $\kappa_x$, $\kappa_y$: the graph of such quantities as a function of $\phi$ do not cross each other, as occurs in panel (\textit{b}) for the theoretical prediction of the isotropic case. The fact that our model accounts for such effect and explains the experimental data when considering the anisotropy of the pdots is a very robust result, showing again that such model, although an approximation, correctly explains the measured data. Therefore, the results of Fig. \ref{anisotropy} confirm the fundamental role of the anisotropy in the dynamics of our pdots.

Finally, it is worth to mention that similar measurements were also performed imposing a circular polarization in our laser. In this case, the measured trap stiffness in the two perpendicular directions ($\kappa_x$, $\kappa_y$) assume values similar to those obtained for $\phi$ = 45$^o$ in each specific case of Fig. \ref{anisotropy}. 

In summary, we presented a novel approach to determine the optical properties of materials in the nanoscale range using optical tweezers (OT), demonstrating that the effective optical properties of our fluorescent pdots can be modulated by changing the light wavelength that excites the sample. Such results open the door for a fine tuning of the pdots optical response, with possible applications in the optoelectronic industry.

This research was funded by Conselho Nacional de Desenvolvimento Científico e Tecnológico (CNPq), Financiadora de Estudos e Projetos (FINEP), Fundação de Amparo à
Pesquisa do Estado de Minas Gerais (FAPEMIG), Coordenação de Aperfeiçoamento de Pessoal de Nível Superior (CAPES) - Finance Code 001, and INCT of Spintronics and
Advanced Magnetic Nanostructures (INCT-SpinNanoMag), CNPq 406836-2022-1. M. S. Rocha acknowledges C. I. L. de Araújo for helpful discussions on the results.

\bibliography{Pdots_bibtex}	
	
\end{document}